# Anomalous Molecular Weight Dependence Reveals the Origin of Mechanical Enhancement in Polymer Nanocomposite

Tae Yeon Kong, So Youn Kim*

School of Chemical and Biological Engineering, Institute of Chemical Processes, Seoul National University; Seoul, 08826, Republic of Korea

KEYWORDS: Polymer nanocomposite, Polymer adsorption, Conformation, Bimodal molecular weight, Interfacial layer, Preferred adsorption

Polymers at the interface exhibit frustrated chain conformation distinct from the bulk, impacting particle dispersion and properties of polymer nanocomposites (PNCs). Utilizing bimodal molecular weight (MW) PNCs, we observed unusual chain elongation of interfacial polymers, which promotes networked structures at low particle loadings. Paradoxically, reducing the average MW in bimodal PNCs significantly enhances the shear modulus by up to $10^3$ times, challenging the conventional preference for higher MW in property enhancement. The mechanical reinforcement, aided by stretched chains, manifests when the short chain is substantially shorter than the long chain ($R_g$ ratio $>\sim 2$), indicating a universal stretching factor of 2.3.

# INTRODUCTION

Polymer nanocomposites (PNCs) are widely used for their superior physical properties achieved by incorporating nanoparticles (NPs) into a polymer matrix.[1-5] Tremendous efforts have been made to identify the origins of their excellence in properties,[6-9] which are believed to stem from the well-defined NPs-polymer interface.[10-13]

When polymers are adsorbed on NPs, interfacial layers are formed, controlling the microstructure of NPs and ultimately determining the physical properties of PNCs.[14-16] The polymers at interfacial layers are responsible for the good dispersion of NPs, and exhibit distinct features in their dynamics and conformation compared to those in bulk. Thus, proper modification of the interfacial layer through NPs and polymer parameters tailors the PNCs with desired properties.[16-20]

Polymer molecular weight (MW) is a crucial parameter in tuning the interfacial layer.[21-23] High MW polymers enhance many physical properties of PNCs,[24] forming thick interfacial layers in PNCs whose thickness corresponds to the radius of gyration, $R_g$.[23,25,26] However, excessively high MW increases viscosity and processing challenges, leading to packing frustration of adsorbed layers[27] and reducing the physical properties of the PNCs.[21] Thus, the design of the PNCs requires careful consideration of the polymer MW to optimize both bulk and interfacial properties.

In this regard, we studied the PNCs with a bimodal MW distribution by blending high and low (unentangled) MW polymers. Poly(ethylene glycol) (PEG) with 0.4 and 10 kg/mol ($M_n$) is employed for the unimodal/bimodal MW matrix (Figure S1). For the bimodal matrix, the PEGs with two MWs are blended at different volume ratios of short polymer ($R_{short}$): $R_{short}$ =0, 1, and 0.5 indicate long unimodal-, short unimodal-, and 50:50 bimodal MWs of the matrices. The PNCs with varying particle volume fractions ($\phi_c$) were produced by adding silica NPs (Figure S2) to the unimodal/bimodal polymers with complete solvent evaporation.

## RESULTS AND DISCUSSION

The physical properties of both polymers and PNCs generally increase with MW in unimodal MW systems.[22] In bimodal MW systems, blending short polymers with long polymers (increasing $R_{short}$) reduces the average MW; thus, one can predict that the yielding properties of both polymers and PNCs are proportional to the average MW similar to unimodal MW systems. As expected, the inset of Figure 1a (and Figure S3a) confirms that the complex modulus (G*) of the neat polymer with bimodal MW systematically decreases with $R_{short}$ as depicted in Figure 1b (black line).

Similarly, with nanoparticles added in bimodal MW matrices, a decline of modulus is expected to follow MW dependency (dotted grey). However, we found bimodal MW PNC exhibits non-monotonic MW behavior (Figure 1a and S3): bimodal PNCs ($R_{short}$=0.5) show remarkably higher G* compared to unimodal PNCs ($R_{short}$=0 and 1) depicted as a multicolor line in Figure 1b. The decrease in average MW rather increases the physical properties of PNCs against the general perception standing in polymer physics.

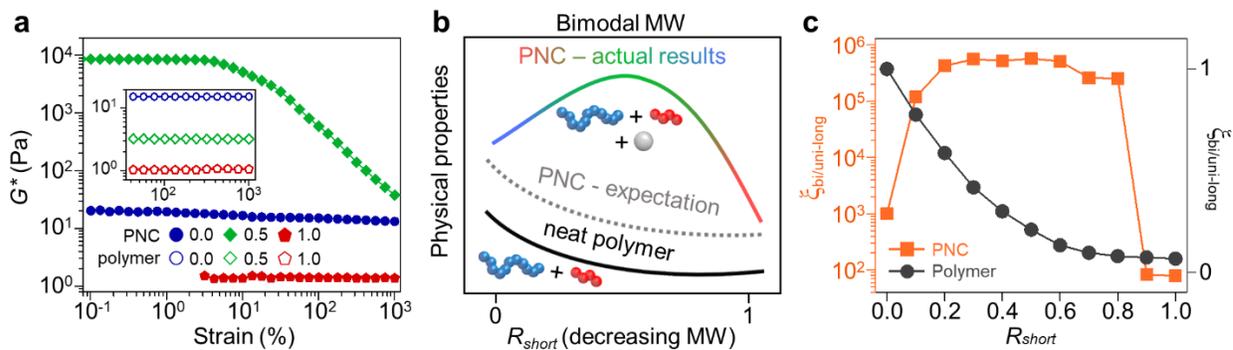

**Figure 1.** (a) Complex shear modulus (G*) obtained with strain sweep experiments (at 1 Hz, 75 °C) for PNCs at $\phi_c$=0.1 and (inset) for neat polymer with varying $R_{short}$. (b) Non-monotonic molecular weight (MW) dependence of physical properties in bimodal PNC. (c) Enhancement factor, $\zeta_{bi/uni\text{-}long}$, of PNC (orange) and neat polymer (black).

For more comparison, the enhancement factor, $\zeta_{bi/uni\text{-}long}$, is defined as the G* of bimodal PNC in the linear viscoelastic region normalized with that of 10k unimodal PNCs ($R_{short}$=0). In Figure 1c, $\zeta_{bi/uni\text{-}long}$ of neat polymer (black) linearly decreases with $R_{short}$ due to average MW reduction. However, $\zeta_{bi/uni\text{-}long}$ of PNC (orange) rapidly increases with $R_{short}$, reaching maximum G* at $R_{short}$= ~0.5, and sharply decreasing at 0.9. This indicates that despite the decrease in the average MW of the matrix, G* is significantly improved in PNCs, rather by several thousand times.

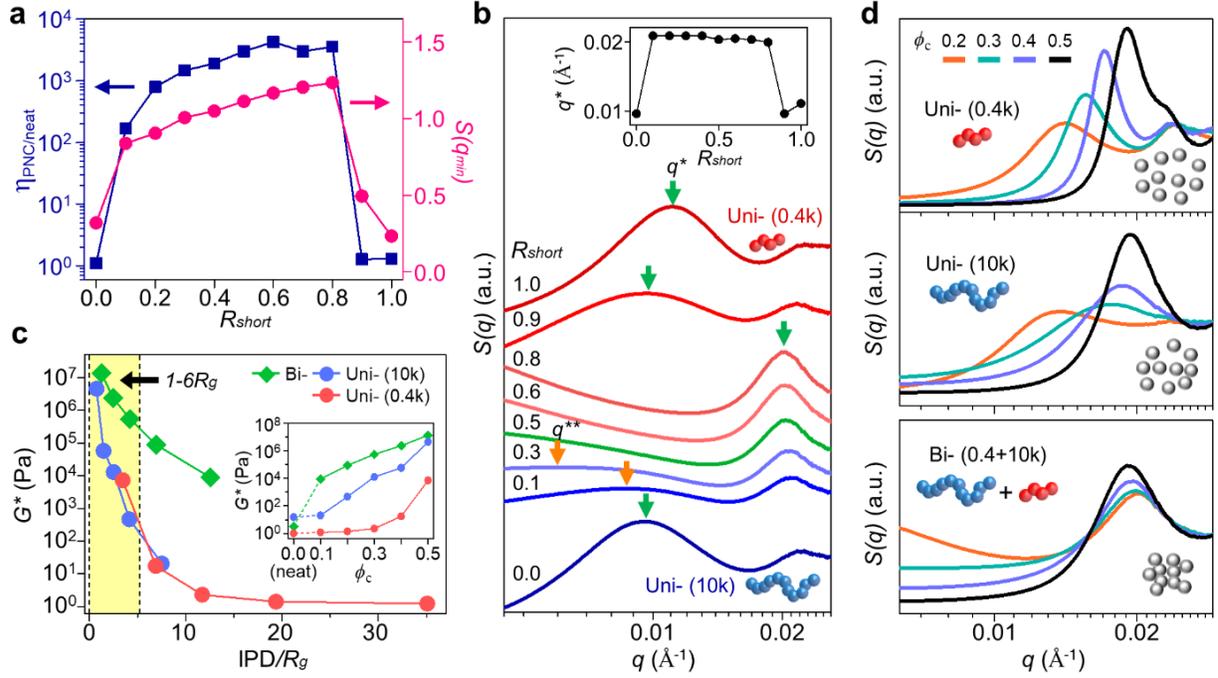

**Figure 2.** (a) Degree of enhancement, $\eta_{PNC/neat}$, and $S(q_{min})$ with varying $R_{short}$ at $\phi_c=0.1$. (b) Structure factor, $S(q)$, for different $R_{short}$ as labeled. (inset) $q^*$ variation with $R_{short}$ at $\phi_c=0.1$. (c) $G^*$ of unimodal and bimodal ($R_{short}=0.5$) PNCs according to the IPD/$R_g$. (inset) Measured $G^*$ with varying $\phi_c$ ($\phi_c = 0.0$ for neat polymers). (d) $S(q)$ of PNCs with varying $\phi_c$ for unimodal (0.4k), unimodal (10k), and bimodal PNCs.

This discrepancy found in bimodal PNC but not in neat polymers suggests that the particle-polymer interfacial layer plays a crucial role. To confirm the role, the degree of enhancement, $\eta_{PNC/neat}$ is defined as $G^*_{PNC}/G^*_{neat}$, and presented in Figure 2a. The $\eta_{PNC/neat}$ further confirmed that replacing long polymers with more short polymers in the bimodal matrix (increasing $R_{short}$) led to increased enhancement, highlighting their significant contribution to interfacial layer modification and mechanical enhancement in bimodal PNCs.

To understand the unusual modulus enhancement, the microstructure of particle dispersions was examined employing small angle x-ray scattering (SAXS) experiments.[28,29] The scattering structure factors with the scattering vector, $q$, are presented in Figure 2b with varying $R_{short}$ from 0.0 to 1.0 at a fixed $\phi_c$=0.1 (see also Figure S4).

One finds the peak position, $q^*$ (green), corresponding to particle correlation, shifts with increasing $R_{short}$. Unimodal PNCs exhibited $q^*$ near 0.01 Å$^{-1}$ indicating uniform dispersion without aggregation (Figure S5). However, $q^*$ of bimodal PNCs is shifted to ~0.02 Å$^{-1}$ ($qD_{NP}$~7) indicating that the particles are in contact except $R_{short}$=0.9. Additional peaks, $q^{**}$ (orange) appear near 0.009 Å$^{-1}$ at $R_{short}$ =0.1, indicating particle clusters, and $q^{**}$ shifts to the lower $q$ with increasing $R_{short}$ implying larger cluster sizes. The $q^{**}$ eventually disappears from the observable range of $q$ above $R_{short}$ = 0.5, and apparent upturns at low $q$ are found, indicating larger agglomeration. We found that this structural change with $R_{short}$ resembles the property change with $R_{short}$. Comparing the $q^*$ in the inset of Figure 2b and the $\zeta_{bi/uni\text{-}long}$ of PNC (orange) in Figure 1c, it is evident that the spatial organization of particles determines the physical properties of PNC. Agglomerated particles in bimodal PNC drive stronger modulus enhancement.

Figure 2a also shows there is an apparent correlation between $S(q)$ at the lowest $q$, $S(q_{min})$, and $\eta_{PNC/neat}$. Noting the $S(q_{min})$ is related to the osmotic incompressibility of the systems and increases as particles are more attractive,[30] this remarkable consistency aligning the $q^*$ variation indicated that the increasing attraction with $R_{short}$ creates larger particle aggregation networks, contributing to higher $\eta_{PNC/neat}$.

This counterintuitive property enhancement by adding short polymers occurs exclusively in the presence of particles. Figure 2c inset and Figure S6 show the $G^*$ (at LVER) of the unimodal and bimodal systems with $\phi_c$. Upon adding particles, all PNC exhibit modulus enhancement with increasing $\phi_c$, accompanied by a decrease in the average interparticle distance (IPD).[31] As IPD decreases, it is known that an interfacial (adsorbed) layer thickness corresponds to $R_g$ begin to interact at a few $R_g$ and overlap at IPD~$2R_g$.[22,32] Consequently, the modulus can be significantly increased as bridging-mediated particle networks are more readily created through the mediation of bulk polymer. Figure 2c finds that the high MW unimodal PNC reaches the range of 1-6 IPD/$R_g$ faster than the low MW unimodal PNC (estimated average IPD of PNCs are presented in Figure S7). Notably, a significant enhancement is found around IPD~$4R_g$ such that $\phi_c$=0.5 for 0.4k ($\phi_c$=0.2 for 10k) unimodal PNCs.

However, the most drastic improvement in mechanical properties occurs with bimodal PNCs, reinforced by 2.5-4 orders at $\phi_c$=0.1 where the IPD of bimodal PNC is expected to be far beyond a few $R_g$, which suggests the creation of the non-conventional-interfacial-layer in bimodal PNC distinct from the classical interfacial layer corresponding to $R_g$.

Figure 2d presents $S(q)$ variation with $\phi_c$ (Figure S8 for $I(q)$). In 0.4k unimodal PNC, well-dispersed particles limit modulus enhancement due to a thin interfacial layer ~1 nm ($R_g$~0.8 nm) that restricts entanglement or bridging. In 10k unimodal PNC, particles are less ordered with $\phi_c$, and bridging occurs at a sufficiently close IPD (<$4R_g$) resulting in particle aggregation and high mechanical properties (Figure 2c), particularly at $\phi_c$~0.4-0.5, where IPD is <$2R_g$.

However, the bimodal PNC microstructure differs qualitatively and quantitatively, not explained by conventional IPD changes. All structure peaks (~0.02 Å$^{-1}$) imply particle contact regardless of $\phi_c$ with $S(q^*)$ reflecting the development of more coherent and robust aggregation networks, which results in more superior mechanical properties than unimodal PNCs.

Thus, while the change of microstructure and the modulus enhancement of unimodal PNC align with the concept of effective IPD, that of bimodal PNC displays distinct characteristics, suggesting a non-conventional interfacial layer.

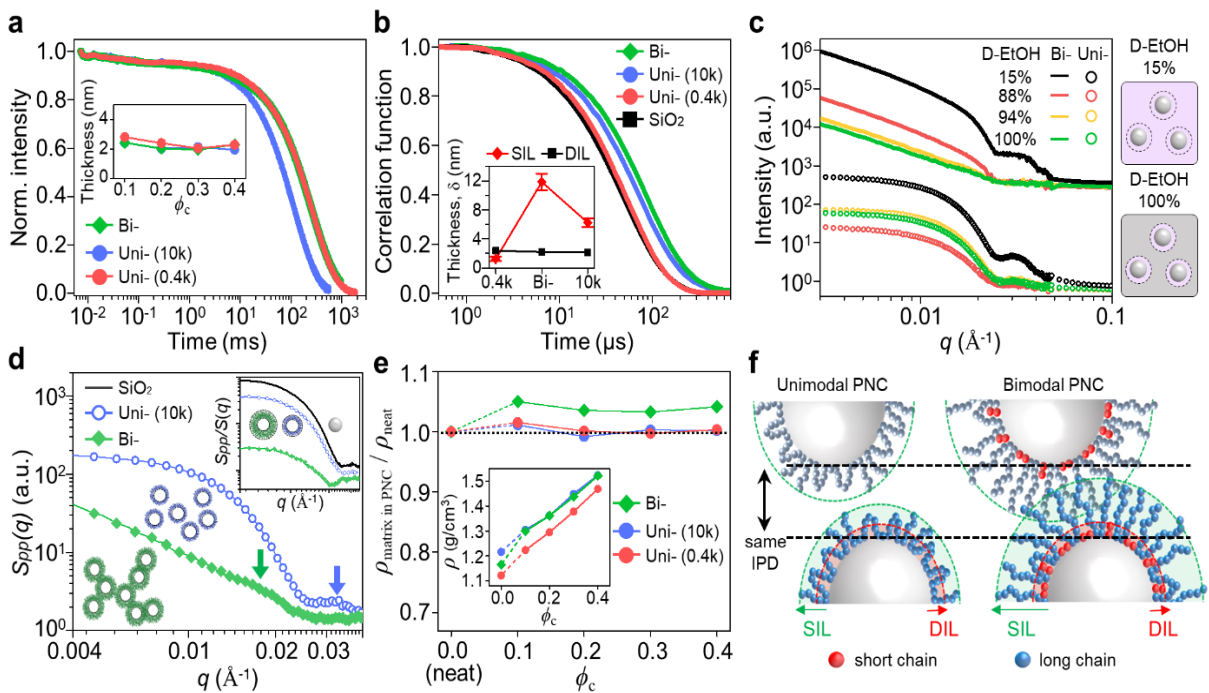

**Figure 3.** (a) Normalized FID and CPMG intensities of PNCs at $\phi_c=0.1$. (inset) interfacial layer thickness with varying $\phi_c$. (b) The correlation function of bare $SiO_2$ NPs and polymer adsorbed NPs measured by DLS. (inset) the averaged thickness of SIL and DIL (c) SANS intensities of concentrated PEG/silica solution with varying deuterated ethanol concentration at 70 °C with illustration of contrast-matched silica and PEG layers. (d) Polymer collective structure factor, $S_{pp}$ for different MW systems. The inset shows the $S_{pp}/S(q)$ and particle form factor comparison. (e) The normalized polymer density, $\rho_{\text{matrix in PNC}}/\rho_{\text{neat}}$, with varying $\phi_c$ (f) Schematic diagram of DIL and SIL in long unimodal and bimodal PNCs.

Two types of interfacial layers are recognized at interfaces: the dynamical interfacial layer (DIL) and the structural interfacial layer (SIL), known as bound polymer layer. While the DIL refers to a region where the dynamics of adsorbed polymers are typically slower by 1~2 orders compared to polymers in bulk,[27,33] the SIL is characterized by adsorbed polymers exhibiting different chain conformation (trains, loops, and tails) from the bulk.[23,33] The thickness of the SIL depends on the polymer MW and can be either thicker or thinner than the DIL.[33]

To characterize the thickness of the DIL, the dynamics of the interfacial polymer were investigated through $^1$H NMR FID and Carr–Purcell–Meiboom–Gill (CPMG) experiments.[16,17,34,35] Figure 3a and Figure S9 compare the full decay signals of the unimodal and bimodal neat polymers and PNCs. While the 10k unimodal PNC shows faster decay in the long-time region (>20 ms) implying the slower relaxation of bulk polymer than 0.4k unimodal and bimodal PNCs, all three relaxations decay similarly in the short-time region. The intensities are globally fitted by three-components stretched exponential function also called the Kohlarusch-Williams-Watts (KWW) function.

The DIL fraction of the polymers is about 0.05 at $\phi_c$=0.1 (Table S1) and the thickness of the DIL was calculated for varying $\phi_c$. Two intriguing findings emerged: first, both unimodal and bimodal PNCs show nearly the same DIL thickness (~2 nm) regardless of MW. Second, despite dramatic changes in the microstructures of unimodal and bimodal PNCs with $\phi_c$, there was no variation in DIL thickness. This consistency aligns with previous reports stating DIL is MW-independent[27,34] and has a thickness of 2-6 nm.[36,37] The DIL thickness was thicker or thinner than the polymer's $R_g$ such that the DIL thickness of unimodal 10k is about half of its $R_g$. This insensitiveness of DIL suggests that DIL has limitations in explaining the difference in structure and properties between unimodal and bimodal PNCs.

To investigate the chain conformation of interfacial polymers regarding the SIL, we utilized dynamic light scattering (DLS)[23] and small angle neutron scattering (SANS).[26,38] Figure 3b shows the correlation functions of NPs with and without polymers. Silica in the 10k decays slower than in the 0.4k implying the formation of a thicker SIL with 10k. However, the bimodal showed the surprisingly thickest SIL with the slowest decay. Considering the theta-condition of the PNC system, the DLS experiment is additionally performed at 60 °C (Figure S9) noting water becomes a theta-solvent near 60 °C.[39,40] Both results show the slowest decay of bimodal systems, confirming the thickest SIL. Figure 3b inset shows SIL thickness varies with MW while the DIL thickness remains constant. In unimodal, the SIL thickness was proportional to $R_g$, with consistent thicknesses of about $1.5R_g$ for both 0.4k and 10k. These results align with previous results where the adsorbed polymer on NPs exhibits a thickness of approximately $1.5R_g$, regardless of whether the polymer is in solution or melt.[22,23,25,31,41,42] However, the bimodal exhibited a surprisingly thick SIL, almost twice that of 10k, suggesting a distinct chain conformation in the bimodal, not proportional to $R_g$.

To directly observe the interfacial layer (SIL) in PNCs, contrast-matching SANS experiments were conducted on a concentrated PEG/silica solution ($\phi_c$=0.1), providing selective information about adsorbed polymer. Experimental details are provided in the Supporting Information.

The scattered neutron intensity $I(q)$ is the sum of three contributions for a given $\phi_c$: $I(q)$ ~ $A\Delta\rho_c^2 P_c(q)S_{cc}(q) + B\Delta\rho_c\Delta\rho_p P_c(q)^{0.5}S_{pc}(q) + C\Delta\rho_p^2 S_{pp}(q)$ where $S_{ij}(q)$ represents structure factors for two components (*pp*, *pc*, *cc*), where subscripts *p* and *c* denote polymer and particles. The $\Delta\rho_j$ is the difference between the scattering length density of component j and the medium. A, B, and C are constants. Figure 3c and S11 show that total scattered intensity, $I(q)$ changes dramatically with systematic $\Delta\rho_j$ variation. At a matched condition, structurally adsorbed polymer layers can be revealed, and all partial structure factors were obtained employing a multiple linear regression fitting method.

In Figure 3d, the extracted polymer partial structure factor, $S_{pp}(q)$, shows a spatial correlation between adsorbed polymer segments. For unimodal MW (10k), $S_{pp}(q)$ resembles particle form factors, indicating density fluctuations of adsorbed polymers on the particle surface. The plateau at low $q$ suggests stable adsorption layer formation corresponding to $R_g$.[38] Conversely, bimodal exhibits a different profile with a shifted hump at 0.017 Å$^{-1}$, and a notable upward trend at lower $q$, indicating larger correlations among interfacial polymers near particles, extending towards the bulk and indicating bridging occurrence.

In the inset of Figure 3d, the normalized $S_{pp}(q)$ with $S(q)$ is presented, which contains partial information on the form factor of the adsorbed polymer shell. The normalized $S_{pp}(q)$ of 10k PNC decreases more rapidly than the particle form factor, $P_c(q)$, indicating the formation of the polymer shell through polymer adsorption. Furthermore, the peak of the bimodal system shifts towards lower $q$ values compared to the unimodal, implying a substantially thicker polymer shell with extended chain conformation in the bimodal PNC. Collectively, DLS and SANS findings strongly confirm a thicker adsorption layer in bimodal PNC with a stretched chain conformation.

Different chain conformations at the interface can yield varied degrees of chain packing, affecting PNC density. Figure 3e shows that bimodal MW neat polymer density falls between two unimodal (inset), resembling rheological properties. However, bimodal PNC density increases more rapidly upon adding particles than unimodal PNCs.

The polymer densities excluding particles ($\rho_{matrix\ in\ PNC}$) are calculated: $\rho_{matrix\ in\ PNC} = (\rho_{PNC} - \rho_{NP}\phi_c)/(1-\phi_c)$, then, normalized with that of neat polymer ($\rho_{neat}$).[21,27] The polymer density in unimodal PNC is almost identical to that of the neat polymer. However, in bimodal PNCs, the polymer density increases by 5%, implying densely packed chain conformations. The thicker and denser SIL layers may result from competitive adsorption between short and long-chain polymers.

When bimodal MW polymers adsorb to a surface, which chain length is preferentially adsorbed is related to the thermodynamic changes due to adsorption. Assuming that the energetic interactions between chains and the nanoparticle surface are the same with chain length due to their chemical identity, the surface becomes enriched with short-chain polymers because the entropic penalty for adsorption is larger for longer chains.[43-45] Additionally, considering that the PEG end -OH group can form hydrogen bonds with silanol groups, the adsorption of shorter chains, which have more end groups per unit volume, may be favored not only entropically but also enthalpically.

In bimodal PNCs, the preferred adsorption of short-chain polymers into long-chain polymers results in a much greater amount of chain adsorption than expected (Figure S12), creating dense interfacial layers as drawn in Figure 3f. This tightly packed adsorption of short chains inhibits the adsorption sites for long-chain segments and facilitates the formation of conformations such as loops and tails to increase the conformational entropy, promoting the stretching of adsorbed long chains and consequently increasing the SIL thickness.[46-48]

Mutual adsorption in bimodal PNC is not due to metastability or non-equilibrium states.[16,47] Changing the order of adding long and short polymers yields consistent microstructures and physical properties in all PNCs (Figure S13).

Overall, the SIL crucially influences the microstructure and physical properties of PNCs, being thickest in bimodal PNCs with chain stretching. The stretched SIL promotes chain interactions, facilitating layer-to-layer or layer-to-bulk interactions at longer IPDs in bimodal PNCs compared to unimodal PNCs (illustrated in Figure 3f). This readily forms networked structures, enhancing mechanical properties as discussed in Figure 2c.

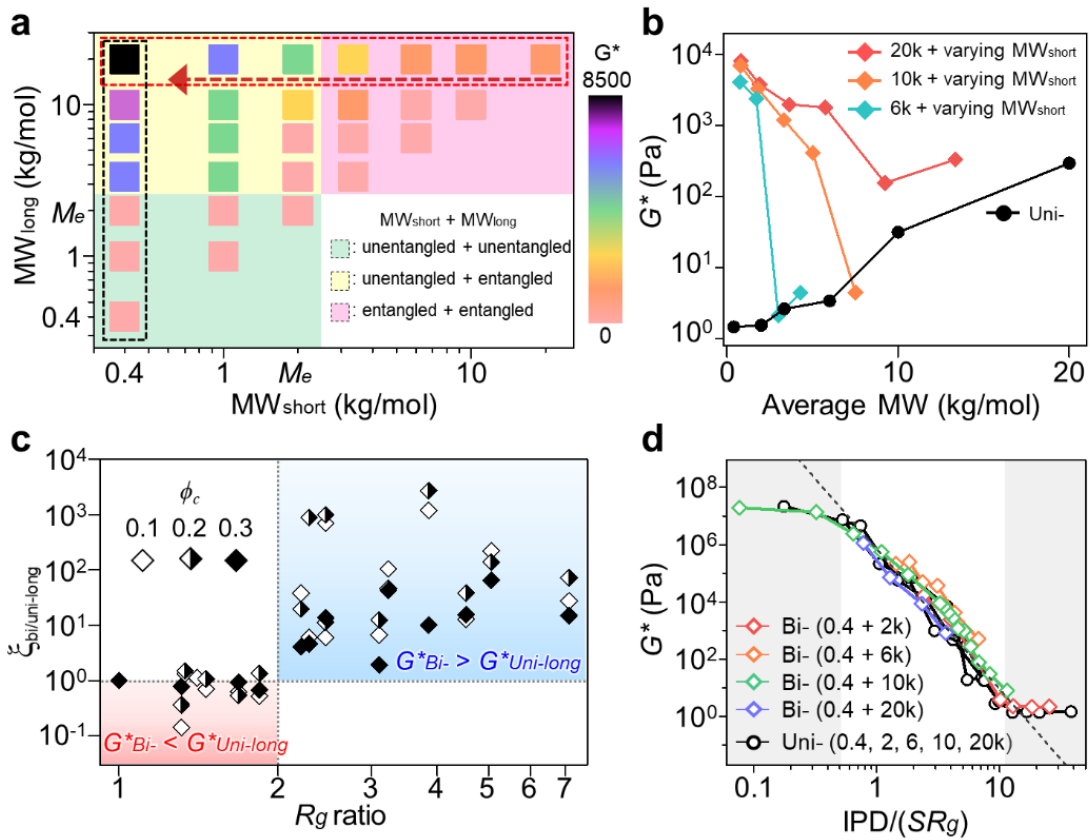

**Figure 4.** (a) $G^*$ of PNCs made with various MW combinations at a fixed $\phi_c$=0.1. ($M_e$~2 kg/mol). The $G^*$ is presented with a color map. (b) $G^*$ of unimodal and bimodal PNCs with average MW at $\phi_c$=0.1. Average MW is calcluuated by using the equation, $MW_{avg}=\sum M_i N_i/\sum N_i$, where $M_i$ and $N_i$ represents the molecular weight and number of molecules with '$i$' repeat units. For the bimodal, $MW_{long}$ is fixed to 6k, 10k, and 20k. (c) $\zeta_{bi/uni\text{-}long}$ according to the $R_g$ ratio ($R_{g,long}/R_{g,short}$). (d) The effective IPD with the stretching factor. The $G^*$ of PNCs according to the IPD/$SR_g$. For the bimodal, $MW_{short}$ is fixed to 0.4k and the $MW_{long}$ changes 2k to 20k. For $R_g$ of bimodal, the $R_g$ of long polymer was used. $R_{short}$=0.5 for (a)-(d).

The entropy penalty for chain adsorption depends on chain length,[43] leading to varying modifications of the SIL based on the choice of two MWs. The modification of the SIL and resulting property enhancement are exploited through the selection of two MWs. Figure 4a investigates the G* of bimodal PNCs across all MW combinations of the $MW_{short}$ and $MW_{long}$.

With a constant $MW_{short}$ (black dashed area), higher $G^*$ values are observed with increasing $MW_{long}$ due to the thicker SIL, more readily engaging in physical interactions with the neighboring SILs or bulk polymers.

The effect of adding short chain MW, $MW_{short}$ is surprisingly counterintuitive on the property enhancement. In a red dotted area, the $G^*$ increases in the left direction as the $MW_{short}$ decreases, which indicates that decreasing the average MW ironically enhances the mechanical property of the PNCs at the same particle loading. This trend is summarized in Figure 4b. Unlike unimodal PNCs, all bimodal PNCs exhibit enhanced G* when the average MW decreases with $MW_{short}$. The decreased entropy penalty for shorter chain adsorption leads to preferred adsorption of shorter chains, altering the chain conformation of long chains and allowing for a larger variation in SIL.

We found that the MW of short polymers is critical; $MW_{short}$ is required to be less than $M_c \sim 2M_e$ for an effective alteration of SIL. In Figure 4a, bimodal PNCs created using two unentangled MWs in the green region show relatively low $G^*$, while those produced with unentangled and entangled MW in the yellow region showed enhanced $G^*$ by more than 3 orders compared to those of unimodal MW PNCs with long-chain. However, if only the entangled MWs (red region) are used, $G^*$ is rather lower than that containing unentangled MW.

There is a critical ratio of the chain lengths between long and short polymers for mechanical enhancement. In Figure 4c, the normalized $G^*$, $\zeta_{bi/uni\text{-}long}$ is explored according to the $R_g$ ratio ($R_{g,long}/R_{g,short}$). When the $R_g$ ratio is less than 2, the $G^*$ of the bimodal PNC does not exhibit significant mechanical enhancement. However, when the short chain is sufficiently short ($R_g$ ratio > 2), all data points fall into the blue region indicating mechanical reinforcement by the inclusion of the short chains. This finding aligns with the results of the Scheutjens-Fleer theory, which suggests that preferred adsorption occurs when the chain lengths differ by a factor of two or more.[49,50]

We further found that the modification of SIL in bimodal PNCs with a critical $R_g$ ratio alters the IPD. The conventional IPD (IPD/$R_g$) concept explains the mechanical enhancement in PNC when particles are sufficiently close within a few $R_g$ allowing for layer-to-layer interactions. The $G^*$ enhancement in unimodal PNCs indeed occurs within the range of IPD/$R_g$ ~1-10 as the SILs interactions are available, creating particle networks. However, in bimodal PNCs, the property enhancement occurs at a much longer IPD~25 (Figure S14).

To accommodate the stretched SIL, a modified IPD concept is proposed, considering a stretching parameter ($S$). Figure 4d shows the $G^*$ of unimodal PNCs (MW=0.4, 2, 6, 10, 20k) according to IPD/$SR_g$, where $S$=1 for unimodal PNC. The degree of chain stretching is characterized by normalizing the IPD of bimodal PNCs with the stretching parameter $S$. Figure 4d shows that $G^*$ curves of all bimodal PNCs superimpose creating a master curve when $S$ is approximately 2.3. The constant of $S$ regardless of the MW$_{long}$ indicates the competitive adsorption of short chains lengthens the SIL by a factor of 2.3, reducing the IPD by the same factor. Note that the SIL thickness of bimodal PNC (Figure 3b) was approximately twice that of unimodal PNC.

We also investigate the modulus enhancement of various bimodal PNCs with different particle sizes, types, and polymers. Figure S15 confirms that property enhancement is a universal phenomenon when particles are incorporated into the bimodal polymer matrix. Noting the particle curvature, number of adsorption sites, and particle-polymer interaction seem to be all different depending on the type of PNCs, the modulus enhancement of bimodal PNC is always larger than that of unimodal PNCs with a few notable changes. As the size of silica NPs decreases, the increased curvature reduces the difference in the thickness of SIL between bimodal and unimodal thus reducing the degree of enhancement.[17] Despite the variation in the particle dimension and the polymer chain architecture, the better property enhancement in bimodal PNCs qualitatively remains. One considers that a subtle change of particle/polymer interface may lead a different degree of property enhancement in bimodal PNCs, which needs to be further addressed with a careful design of the interface.

## CONCLUSION

In conclusion, we reported anomalous behavior in bimodal MW PNCs and investigated the origin of their non-monotonic property enhancement. The traditional rule of proportional physical properties based on average MW is challenged, as our study reveals that decreasing average MW strengthens PNC properties. This unpredicted outcome is attributed to the preferred adsorption of short-chain polymers, leading to stretchable SIL and networked structures. Effective chain stretching occurs when the short chains are sufficiently short ($R_g$ ratio > ~2) with a universal stretching factor of 2.3. These findings extend beyond PNCs, offering insights into various interfaces where polymer adsorbs.

# EXPERIMENTAL METHODS

**Preparation of Polymer Nanocomposite (PNC)**

PNCs were prepared with poly(ethylene glycol) (PEG) and silica nanoparticles (NPs). PEG with seven different MWs, 0.4, 1, 2, 3.35, 6, 10, 20 kg/mol (corresponding $R_g$ is 0.83, 1.3, 1.9, 2.4, 3.2, 4.2, 5.9 nm, respectively), were purchased from Sigma Aldrich and used intactly. Silica NPs with a diameter of 38 nm were synthesized by the Stöber method[30] and used as dispersed in ethanol. The weight fraction of the NPs solution was about 10 wt%. The average size of silica NPs was determined by dynamic light scattering (DLS) and by fitting small angle x-ray scattering (SAXS) intensity as a spherical form factor. The proper amount of silica NPs solution was added into the PEG with unimodal and bimodal molecular weight (MW) distribution to achieve the desired particle volume fraction, $\phi_c$ (0.1 to 0.5) of PNCs. Then, the solution was vigorously vortexed for a few minutes. The mixed solutions were annealed at 70 °C for 4 days in a vacuum oven to fully evaporate the solvent to obtain PNCs.

**Rheology**

Rheological properties for neat polymers and PNCs were measured by using an MCR302 (Anton Paar) rheometer with the cone-and-plate (CP20-4, diameter, 20 mm; cone angle, 4°) or plate-plate (PP08, diameter, 8 mm) geometries. All samples were measured after thermal stabilization for at least 10 min. PEG and PPG-based PNCs are measured at 75 °C and PVP-based PNCs are measured at 160 °C. The strain sweep experiments were conducted at a 1 Hz frequency.

**Small Angle X-ray Scattering (SAXS)**

SAXS experiments were performed at the 4C SAXS II beamline of the Pohang Accelerate Laboratory (PAL) to examine the spatial organization of the silica NPs in the PNCs. The samples were stabilized at 75 °C for at least 10 min after sample loading. The obtained $q$ range was $0.0045 < q < 0.2$ Å$^{-1}$. The scattered intensity, $I(q)$ can be written as, $I(q)=\phi_c V_c \Delta\rho^2 P(q)S(q) + B$, where $V_c$ is the particle volume, $\Delta\rho$ is the difference of electron scattering length density between silica NPs and PEG matrix, $P(q)$ is the form factor of silica NPs as shown in Figure S2, $S(q)$ is the structure factor of silica NPs, and $B$ is the background. Due to the large $\Delta\rho$, the scattered intensity of PNCs is dominated by silica NPs after subtracting the background intensity, $B$. The $S(q)$ was obtained by dividing $I(q)$ by the $P(q)$.

**$^1$H Nuclear Magnetic Resonance (NMR)**

NMR experiments were performed with a Bruker minispec mq20. The samples were thermally stabilized at least 10 min after sample loading. The free induction decay (FID) and Carr-Purcell-Meiboom-Gill (CPMG) sequences are applied to the samples for analyzing the dynamics of the polymer. The FID, which measures the intensity at the transverse plane after a $\pi/2$ pulse, was used for a short time region ($t < 0.1$ ms) because of the inhomogeneity of the magnetic field. To avoid the magnetic field inhomogeneity, the CPMG experiment is additionally performed to observe a longer time region ($t > 0.1$ ms).

The protons in adsorbed (quasi-rigid) and bulk (mobile) polymers can be distinguished based on their different transverse relaxation times ($T_2$ relaxation). The quasi-rigid polymers have strong dipole-dipole couplings between the protons resulting in rapid decay in free induction decay (FID) with a short $T_2$ in the 20 µs range. On the other hand, the dynamically mobile polymers yield slow relaxation with a long $T_2$.

The whole intensities consisting of FID and CPMG are globally fitted by three components stretched exponential function also called the Kohlarusch-Williams-Watts (KWW) function which consists of quasi-rigid, intermediate, and mobile polymer dynamics, defined as I = $A \sum_{r,i,m} \{f_j \exp[-(t/\tau_j)^{b_j}]\}$, where $A$ is a scaling factor, $\tau$ is the $T_2$ relaxation time, $b$ is the stretching exponent, $f$ is the fraction of polymer segments, and subscripts $r$, $i$, and $m$ indicate quasi-rigid, intermediate, and mobile segments, respectively. The intermediate phase is introduced to assess the gradual mobility changes from the particle surface to the bulk. The thickness of the dynamical interfacial layer ($\delta_{DIL}$) is calculated as follows: $\delta_{DIL} = R_{NP}[\{(f_r+f_i)/\phi_c+1\}^{1/3}-1]$.

**Dynamic Light Scattering (DLS)**

DLS experiments were performed with a Zetasizer Nano ZS90 (Malvern) to obtain the SIL thickness at 60 and 70 °C for ethanol and water, respectively. We prepared dilute solutions for polymer and silica NPs (PEG/silica NPs/ethanol or PEG/silica NPs/water) to avoid any interactions between particles. The weight fraction of silica NPs was fixed at 0.1 wt% for all solutions. Each PEG concentration is fixed at 1.7 g/L. We use the viscosity of solvent as the viscosity of the solution due to the dilute condition. The hydrodynamic diameters for bare silica NP and PEG adsorbed silica NP were obtained by fitting the auto-correlation function. The SIL thicknesses were obtained by subtracting the diameter of bare silica NP from the diameter of PEG-adsorbed silica NP and dividing the result by a factor of 2. The detailed experimental procedure to obtain the SIL thickness can be also found in the reference.[23]

**Small Angle Neutron Scattering (SANS)**

SANS experiments were performed using the 40 m SANS instrument at HANARO of the Korean Atomic Energy Research Institute (KAERI) in Daejeon, Republic of Korea. Two sample-to-detector distances of 1.16 and 17.5 m were used to cover the q range of $0.004 < q < 0.4$ Å$^{-1}$. The scattered intensity was corrected for background, empty cell scattering, and the sensitivity of individual detector pixels. The corrected data sets were placed on an absolute scale through the secondary standard method. The samples were loaded into 1 mm path-length quartz cell. The cell temperature was maintained at 70 °C.

**Density measurement**

The densities of polymers and PNCs were measured with a gas displacement pycnometer, AccuPyc II. Each sample was measured at least 9 times and the average density was used.

**Thermogravimetric analysis (TGA) procedure**

TGA was used to measure the amount of the adsorbed polymer onto the silica NP surface. We first prepared the same solutions as when we created the unimodal and bimodal PNCs with $\phi_c$=0.1. To remove the free polymers that were not adsorbed onto the NPs, the solutions were centrifuged at 17,000 rpm for 90 mins. Then supernatant containing free polymer was removed and the precipitate consisting of the NPs in which polymers are adsorbed was redispersed in ethanol by ultrasonication. For the complete removal of the free polymer, this process was performed at least three times. The final precipitate was dried in a vacuum oven for 24 hours before being used in TGA. The dried samples are burned in a TGA with a sequence as follows: isothermal at 80 °C for 10 minutes and temperature increase from 30 to 800 °C at 10 °C/min. We did not consider the influence of silica in comparison because we used the same silica for all samples.

**The universality of mechanical enhancement in bimodal PNC**

For the preparation of PNCs with various NPs sizes, types of particles, and polymers in Figure S15, silica NPs with diameters of 13 nm and 72 nm were synthesized by the Stöber method. The average sizes of silica NPs were determined by DLS. Poly(propylene glycol) (PPG) with 0.425 and 4 kg/mol were purchased from Sigma Aldrich and used intactly. Poly(vinylpyrrolidone) (PVP) with 10 and 360 kg/mol were purchased from Scientific Polymer Products Inc. and used intactly. Cellulose nanocrystal (CNC) was purchased from Celluforce (NCV100) and used as a dispersed in deionized water. Graphene oxide (GO) was supplied from Standard Graphene. Zinc oxide (ZnO) and alumina ($Al_2O_3$) nanoparticles are purchased from Sigma Aldrich and used as dispersed in deionized water. PNCs were prepared in the same procedure as the previously described PEG/silica PNC. $\phi_c$ is 0.1 for PEG/silica and PVP/silica PNCs and $\phi_c$ is 0.2 for PPG/silica PNC. In the case of PNCs composed of PEG/GO, PEG/CNC, PEG/ZnO, and PEG/$Al_2O_3$ particles were included in the final PNC by 0.5, 1, 15, and 15 wt%, respectively.

## AUTHOR INFORMATION

**Corresponding Authors**

So Youn Kim – School of Chemical and Biological Engineering, Institute of Chemical Processes, Seoul National University, Seoul 08826, Republic of Korea; Email: soyounkim@snu.ac.kr

**Author**

Tae Yeon Kong – School of Chemical and Biological Engineering, Institute of Chemical Processes, Seoul National University, Seoul 08826, Republic of Korea

**Author Contributions**

S.Y.K and T.Y.K initiated the project. T.Y.K. performed all experiments and analyzed the data. S.Y.K supervised the project and analyzed the data. All authors contributed to interpreting the results and preparing the manuscripts.

**Notes**

The authors declare no competing financial interest.

## ACKNOWLEDGMENTS

This work was supported by the National Research Foundation of Korea (NRF) grant funded by Korea Government (MSIT) (NRF-2021R1A2C2007339 and NRF-2021R1C1C2012905). SAXS measurements were performed at the PLS-II 4C SAXS II beamline of the Pohang Accelerator Laboratory (PAL) in Pohang, Korea. SANS experiments were performed using the 40 m SANS instrument at HANARO of the Korean Atomic Energy Research Institute (KAERI) in Daejeon, Republic of Korea.

# Supporting Information

# Anomalous Molecular Weight Dependence Reveals the Origin of Mechanical Enhancement in Polymer Nanocomposite

Tae Yeon Kong, So Youn Kim*

School of Chemical and Biological Engineering, Institute of Chemical Processes, Seoul National University; Seoul, 08826, Republic of Korea

## 1. Unimodal and bimodal MW distribution system.

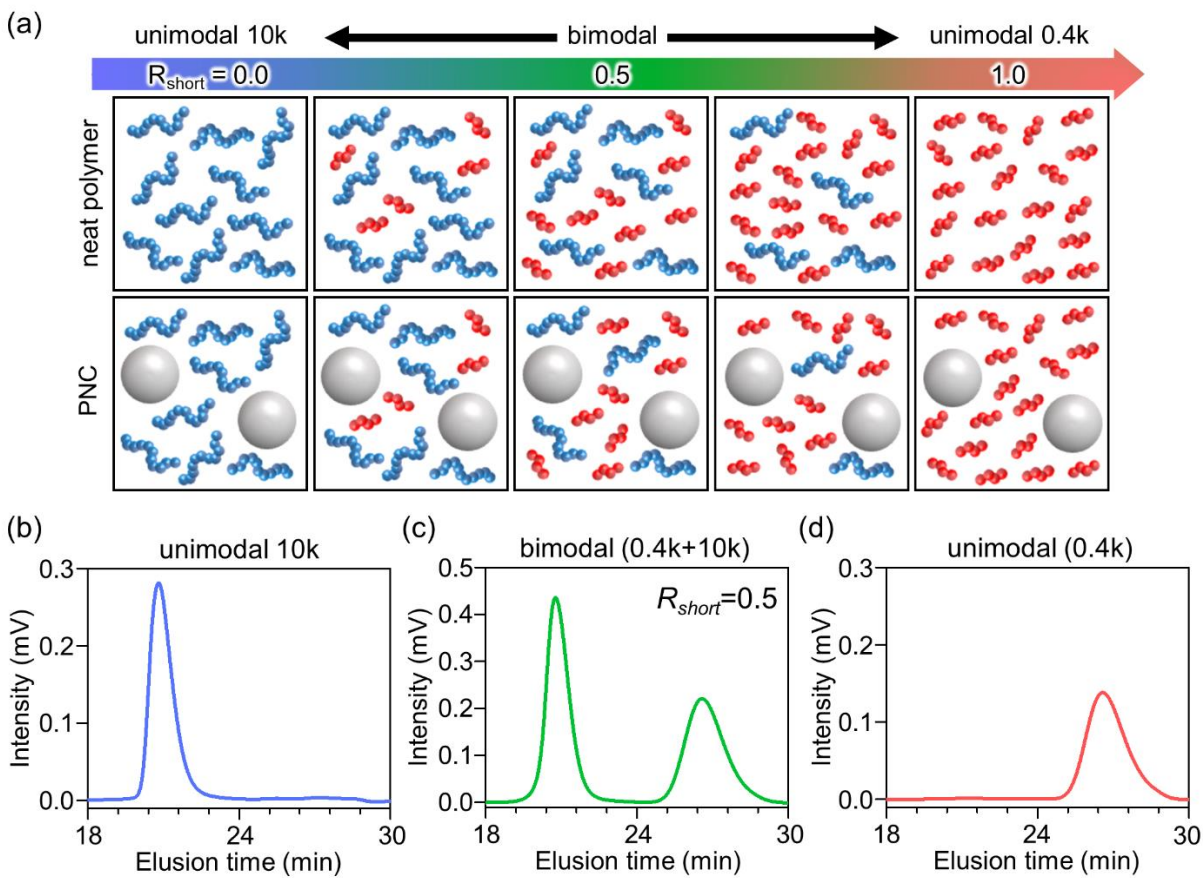

Figure S1. (a) Schematic diagram of the unimodal and bimodal molecular weight distribution systems in neat polymer and PNC. $R_{short}$ represents the volume fraction of peg 0.4k (short polymer) in the polymer matrix. Gel chromatography (GPC) results for polymers with (b) unimodal (PEG 10k), (c) bimodal, and (d) unimodal (PEG 0.4k) MW distribution dissolved in tetrahydrofuran (THF). For bimodal, $R_{short}$ is 0.5. GPC was conducted using an Agilent 1200S. Polystyrene standards were used for calibration and THF was used as an eluting solvent.

2. **Form factor of silica nanoparticles**

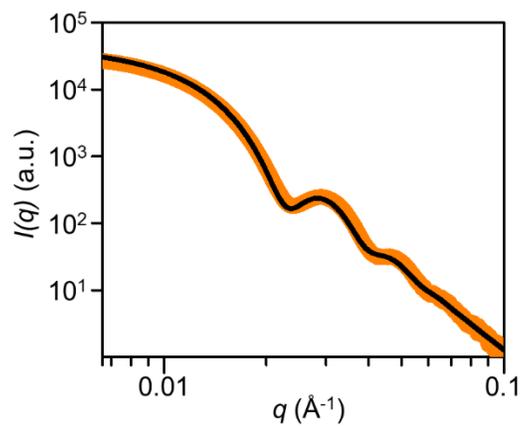

Figure S2. The form factor, *P(q)*, for 38 nm silica nanoparticle (orange). The black curve is the fitted spherical nanoparticle model for *P(q)*. The particle size obtained through the form factor fitting was 38.5 ± 4.2 nm.

## 3. Non-monotonic MW behavior in bimodal MW PNC.

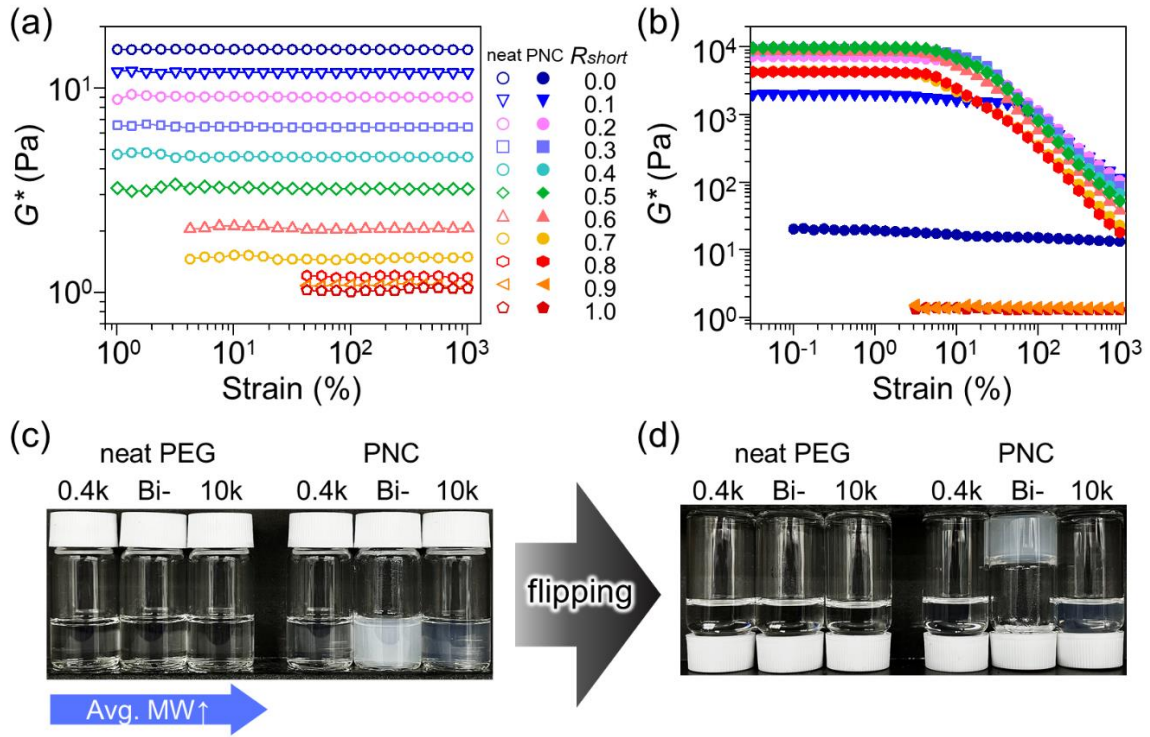

Figure S3. Complex shear modulus ($G^*$) obtained with strain sweep experiments (at 1 Hz) for (a) neat polymer and (b) PNCs at $\phi_c$ =0.1 with varying $R_{short}$. Optical images for neat PEG melt (left) and PNC melt (right) (c) before and (d) after flipping at 70 °C.

## 4. Small Angle X-ray scattering (SAXS) data with varying $R_{short}$

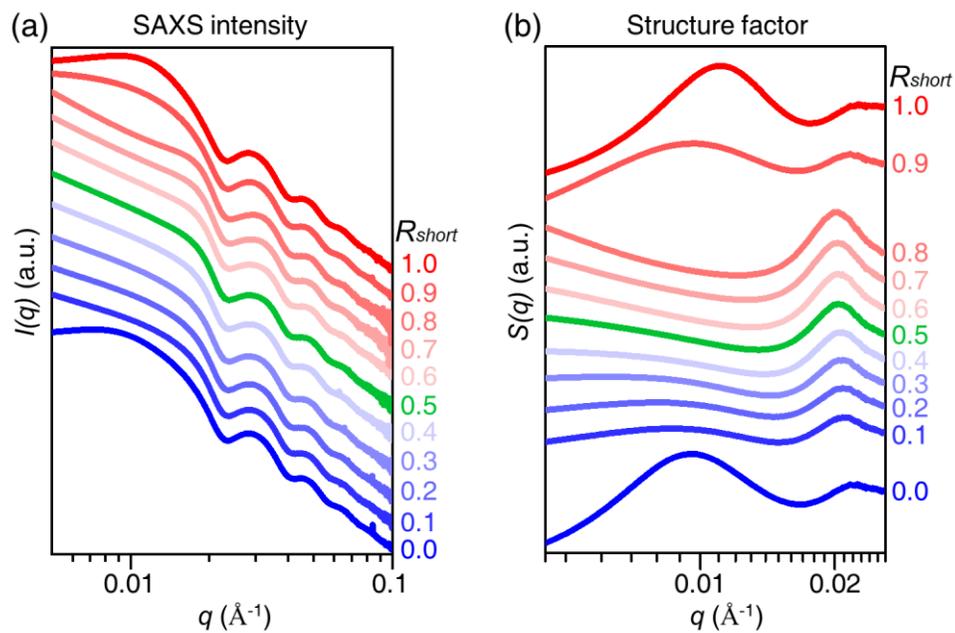

Figure S4. (a) The scattered intensities, $I(q)$, and (b) the structure factors, $S(q)$, of PNCs with varying $R_{short}$ at a fixed $\phi_c = 0.1$. The numbers (0.0 to 1.0) indicate the $R_{short}$ of each profile. The profiles were shifted vertically to be distinguished.

5. **SEM image of unimodal and bimodal PNCs**

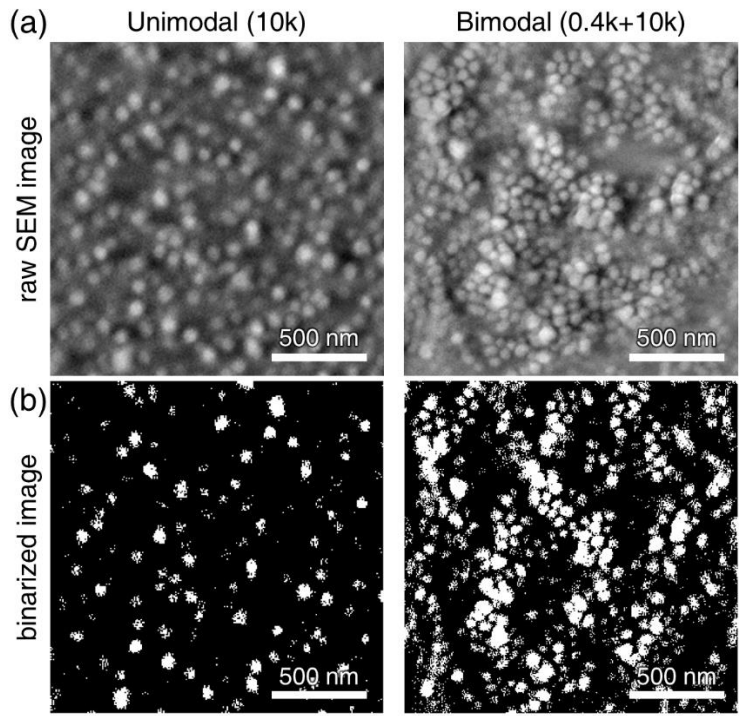

Figure S5. (a) Raw and (b) binarized SEM images for unimodal (left) and bimodal (right) PNC at $R_{short} = 0.5$ and $\phi_c = 0.1$. Carl Zeiss Supra 55VP field-emission scanning electron microscope (high vacuum, 5 keV) was used for imaging.

## 6. Strain sweep results of PNCs

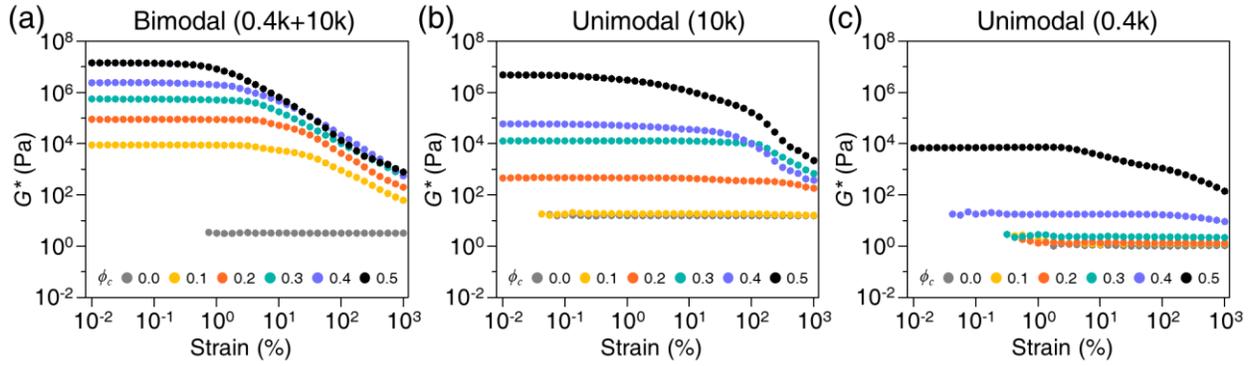

Figure S6. Results of strain sweep experiment for (a) bimodal, (b) unimodal 10k, and (c) unimodal 0.4k neat polymers and PNCs with varying $\phi_c$. All experiments were performed at 1 Hz frequency. For bimodal, $R_{short}$ is fixed at 0.5.

## 7. Interparticle distance of silica NPs with molecular weight

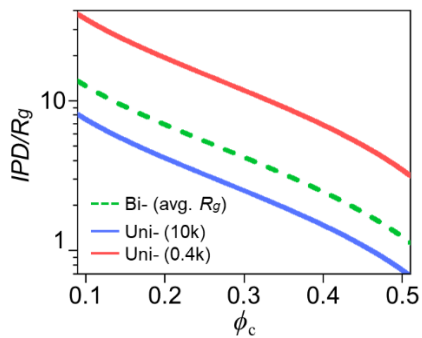

Figure S7. The IPD/$R_g$ variation of unimodal and bimodal PNCs according to the $\phi_c$. Average IPD of silica NPs is estimated as follows: $D_{NP}[\{2/(\pi\phi_c)\}^{1/3}-1]$, where $D_{NP}$ is a particle diameter. The averaged $R_g$ for 0.4k and 10k is used for the $R_g$ of bimodal PNC.

## 8. Small Angle X-ray scattering (SAXS) data with varying $\phi_c$

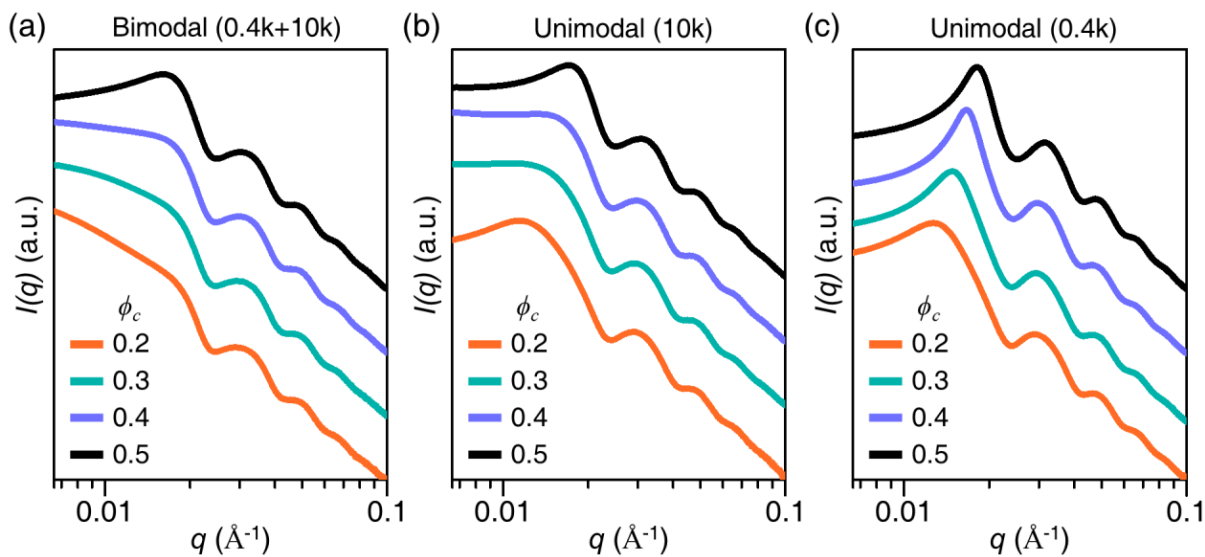

Figure S8. The scattered intensities, *I(q)*, of PNCs with (a) bimodal, (b) unimodal 10k, and (c) unimodal 0.4k with varying $\phi_c$. For bimodal, $R_{short}$ is fixed at 0.5. The profiles were shifted vertically to be distinguished.

## 9. $^1$H-NMR (FID & CPMG) experimental results

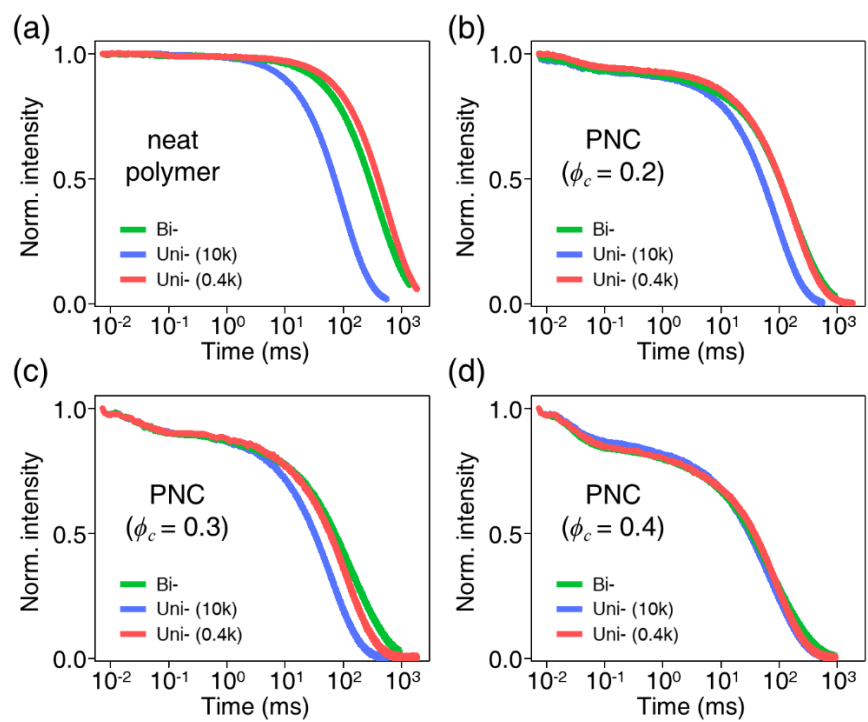

Figure S9. Normalized FID and CPMG intensities of (a) neat polymers, PNCs with (b) $\phi_c = 0.2$, (c) $\phi_c = 0.3$, (d) $\phi_c = 0.4$.

Table S1. $^1$H-NMR KWW fitting results of $T_2$ relaxation times ($\tau_{2,m}$, $\tau_{2,i}$, and $\tau_{2,r}$), polymer fraction ($f_m$, $f_i$, and $f_r$), and thickness of DIL ($\delta_{DIL}$). Subscripts *m*, *i*, and *r* indicates the mobile, intermediate, and quasi-rigid.

| | $\phi_c$ | $\tau_{2,m}$ (ms) | $\tau_{2,i}$ (ms) | $\tau_{2,r}$ (ms) | $f_m$ | $f_i$ | $f_r$ | $\delta_{DIL}$ |
|---|---|---|---|---|---|---|---|---|
| Unimodal 0.4k | neat | 586.3 | | | | | | |
| | 0.1 | 258.1 | 0.200 | 0.0354 | 0.949 | 0.0300 | 0.0212 | 2.804 |
| | 0.2 | 177.3 | 0.121 | 0.0231 | 0.915 | 0.0417 | 0.0438 | 2.393 |
| | 0.3 | 113.9 | 0.101 | 0.0291 | 0.892 | 0.0449 | 0.0630 | 2.049 |
| | 0.4 | 89.89 | 0.0789 | 0.0325 | 0.838 | 0.0747 | 0.0875 | 2.283 |
| Bimodal | neat | 422.5 | | | | | | |
| | 0.1 | 256.2 | 0.200 | 0.0370 | 0.957 | 0.0232 | 0.0202 | 2.426 |
| | 0.2 | 184.4 | 0.125 | 0.0278 | 0.928 | 0.0324 | 0.0392 | 2.039 |
| | 0.3 | 152.1 | 0.0930 | 0.0266 | 0.898 | 0.0481 | 0.0543 | 1.955 |
| | 0.4 | 87.63 | 0.0644 | 0.0269 | 0.837 | 0.0749 | 0.0879 | 2.291 |
| Unimodal 10k | neat | 102.7 | | | | | | |
| | 0.1 | 115.7 | 0.150 | 0.0384 | 0.957 | 0.0227 | 0.0207 | 2.423 |
| | 0.2 | 89.18 | 0.109 | 0.0236 | 0.929 | 0.0249 | 0.0459 | 2.020 |
| | 0.3 | 61.05 | 0.0878 | 0.0281 | 0.888 | 0.0537 | 0.0587 | 2.126 |
| | 0.4 | 72.86 | 0.0541 | 0.0282 | 0.864 | 0.0667 | 0.0692 | 1.945 |

## 10. Dynamic Light Scattering for Structural Interfacial Layer (SIL) at theta condition

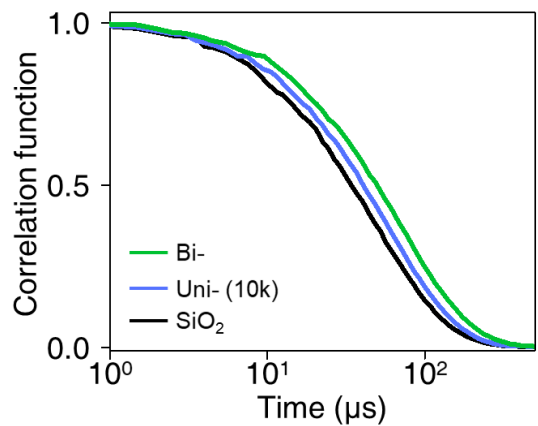

Figure S10. The correlation function of bare $SiO_2$ NPs and polymer adsorbed NPs dispersed in 60 °C water measured by DLS. $R_{short}$ is 0.5 for bimodal.

## 11. Small Angle Neutron Scattering (SANS) results

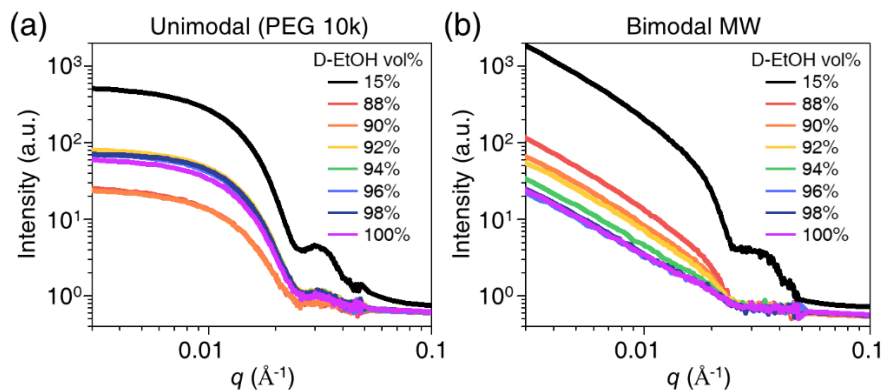

Figure S11. Scattered neutron intensity profiles of (a) unimodal (10k) and (b) bimodal MW PEG/silica concentrated solutions with varying deuterated ethanol (D-EtOH) ratios. Particle volume fraction ($\phi_c$) is fixed at 0.1 and $R_p$ is fixed at 0.4.

For contrast matching, PNCs were redispersed in a mixture of deuterated ethanol ($D_6$-EtOH) and hydrogenated ethanol (H-EtOH). To ensure the highly concentrated systems for PNCs, the volume ratio of the polymer to polymer solution was fixed to 0.4. Scattered neutron intensity profiles for unimodal (10k) and bimodal systems are shown in Figure 3c and Figure S11.

The intensity, denoted as $I(q)$, of scattered neutrons with a wave vector $q$ is the sum of three contributions for a given $\phi_c$: $I(q) \sim A\Delta\rho_c^2 P_c(q) S_{cc}(q) + B\Delta\rho_c\Delta\rho_p P_c(q)^{0.5} S_{pc}(q) + C\Delta\rho_p^2 S_{pp}(q)$ where $S_{ij}(q)$ represents the structure factors associated with the two components (pp, pc, cc), where subscripts p and c denote polymer segments and particles, respectively. The $\Delta\rho_j$ is the difference between the Scattering Length Density (SLD) of component j and the medium. A, B, and C are constants. To determine all three $S_{ij}(q)$ at a fixed $\phi_c$, $\rho_j$ is expressed as a function of the D/H-EtOH ratio in the solvent phase, and the $\Delta\rho_j$ values are altered by appropriately adjusting the D/H-EtOH ratio. Solving these simultaneous equations involves multiple linear regression fitting methods using $I(q)$ measured at least three D/H-EtOH ratios close to $\Delta\rho_c \sim 0$. This process yields $S_{pp}(q)$ and $S_{pc}(q)$. Detailed procedures can be found in the Supplementary Information and related literature [3-5].

Under the condition of $\Delta\rho_c=0$, the scattered intensity arises from non-bulk-like polymer concentration fluctuations associated with adsorbed polymers around the particles. consequently, the intensity becomes as follows, $I(q) \sim S_{pp}(q) \sim S_{pp}*(q)P_s(q)$, where $P_s(q)$ is the polymer-shell form factor as $\phi_c \to 0$, and $S_{pp}*(q)$ is the structure factor associated with correlations between the centers-of-mass (CM) of the polymer shell. We posit that the adsorbed polymer shell CM will, to a good approximation, mirror those experienced by the CM of the nanoparticles ($S_{pp}*(q) \sim S_{cc}(q)$) such that when $\Delta\rho_c = 0$, $I(q) \sim S_{pp}(q) \sim S_{cc}(q) P_s(q)$[3,4].

The inset of Figure 3d shows the $S_{pp}(q)/S(q)$, $P_s(q)$, for both the bimodal and unimodal (10k) systems. In the presence of PEG, the $P_s(q)$ decreases more rapidly than the particle form factor, $P_c(q)$, indicating the formation of the polymer shell through polymer adsorption. Furthermore, the peak of the bimodal system shifts towards lower q values compared to the unimodal, implying the creation of a substantially thicker polymer shell in the bimodal PNC. Collectively, the findings from DLS and SANS strongly confirm that a thicker adsorption layer is established in the bimodal PNC compared to the unimodal PNC with long polymer.

## 12. Thermogravimetric Analysis (TGA) data

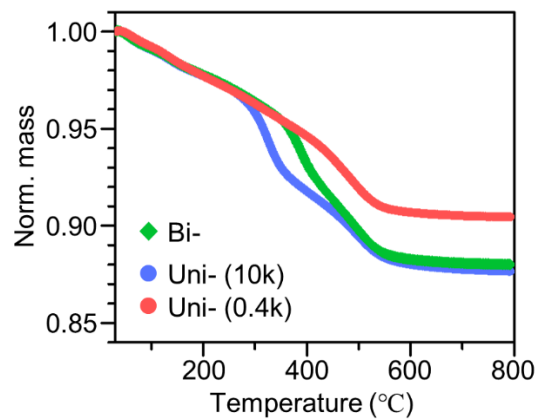

Figure S12. Thermogravimetric analysis (TGA) results of silica NPs adsorbed with bimodal (green), unimodal 10k (blue), and unimodal 0.4k (red) PEG. $R_{short}$ is 0.5 for bimodal.

## 13. Effect of the mixing order on SAXS and rheology of bimodal PNCs

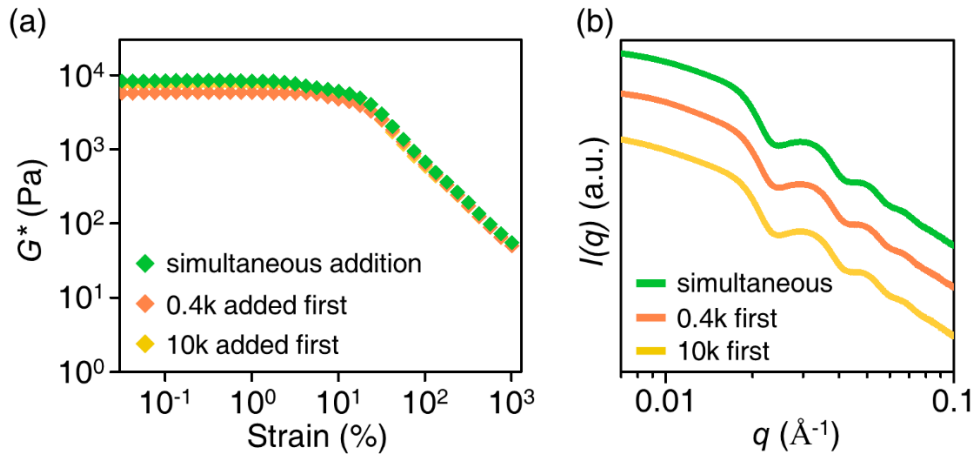

Figure S13. (a) Strain sweep experiments and (b) scattered intensities, $I(q)$, for bimodal PNC ($R_{short}$ = 0.5 and $\phi_c$ = 0.1) with different mixing orders. To confirm whether the bimodal PNC was in equilibrium with this favorable adsorption, we prepared three PNCs: The first PNC was prepared by adding 0.4k and 10k PEG to the silica NPs simultaneously (green) (the) original experimental method). The other PNCs are prepared by adding either 0.4k (orange) or 10k (yellow) PEG first. The first MW of PEG was added to the silica NPs in order to selectively adsorb polymers with that MW and the other MW of PEG was added after 24 hours.

## 14. The change of G* according to IPD/$R_g$ of unimodal and bimodal PNCs

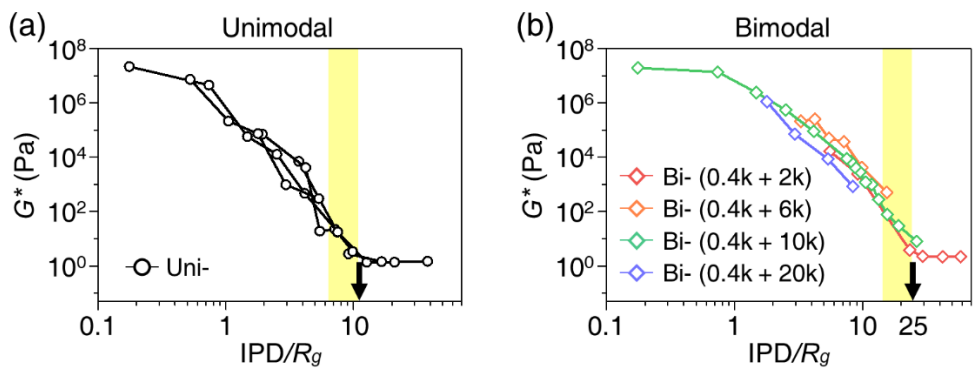

Figure S14. $G^*$ of (a) unimodal and (b) bimodal PNCs according to the IPD/$R_g$. MW$_{short}$ is fixed to 0.4k and the MW$_{long}$ changes 2k to 20k for bimodal PNCs. While the enhancement of $G^*$ begins at IPD/$R_g$~10 for unimodal, that of bimodal begins at a much longer IPD/$R_g$~25.

## 15. Universality for mechanical enhancement in various conditions

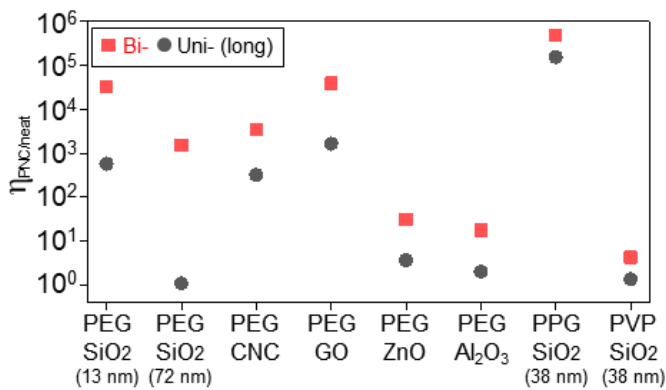

Figure S15. Universality of the higher $\eta_{PNC/neat}$ in bimodal PNCs with varying particle size, types of particles, and polymers. The parentheses indicate the diameter of silica NPs. Experimental details are given in the Experimental section.

Noting the particle curvature, number of adsorption sites, and particle-polymer interaction seem to be all different depending on the type of PNCs, the $\eta_{PNC/neat}$ of bimodal PNC is always larger than that of unimodal PNCs with a few notable changes. As the size of silica NPs decreases, the increased curvature reduces the difference in the thickness of SIL between bimodal and unimodal thus reducing the degree of enhancement. Despite the variation in the particle dimension and the polymer chain architecture, the better property enhancement in bimodal PNCs qualitatively remains. One considers that a subtle change of particle/polymer interface may lead a different degree of property enhancement in bimodal PNCs, which needs to be further addressed with a careful design of the interface.

## 16. Small Angle Neutron Scattering: Determination of partial structure factors

The intensity, *I(q)*, of scattered neutrons at wave vector *q*, has three contributions:

$$I(q) \sim A\Delta\rho_c^2 P_c(q) S_{cc}(q) + B\Delta\rho_c \Delta\rho_p P_c(q)^{0.5} S_{pc}(q) + C\Delta\rho_p^2 S_{pp}(q)$$

where $S_{ij}(q)$ are the structure factors associated with the two components (*pp*, *pc*, *cc*) where the subscripts *p* and *c* indicate polymer segments and particles, respectively; $\Delta\rho_j$ is the difference between the SLD of component *j* and the medium; *A*, *B*, and *C* are constants.

To achieve the contrast matching method, the scattering length density (SLD, $\rho$) of nanoparticles, polymer segments and solvent in highly concentrated polymer/particle solutions must be determined. While the cross sections of silica nanoparticles and PEG are fixed, the contrast relative to the background can be tuned by varying the D/H-EtOH ratio in the solvent phase as shown in Figure S16. In concentrated polymer solutions, the solvent cross section reflects its composition and is written: $\rho_s = \phi_h \rho_h + \phi_d \rho_d + \phi_p \rho_p$, where $\rho_h$ and $\rho_d$ are the cross sections of H-EtOH and D$_6$-EtOH, respectively, while $\rho_p$ is the polymer scattering cross section. Here, $\phi_h$, $\phi_d$ and $\phi_p$ represent the mass fractions of H-EtOH, D$_6$-EtOH and polymer in the continuous phase, respectively, such that $\phi_h + \phi_d + \phi_p = 1$. We fixed $\phi_p = 0.4$ ($R_P = 0.4$) and varied $\phi_d/(\phi_h + \phi_d)$. Our initial studies established $\rho_c = 3.74 \times 10^{-6}$ Å$^{-2}$, $\rho_p = 6.60 \times 10^{-7}$ Å$^{-2}$, $\rho_h = -0.35 \times 10^{-6}$ Å$^{-2}$, $\rho_d = 6.16 \times 10^{-6}$ Å$^{-2}$ such that the contrast match condition for silica is achieved at $\phi_d/(\phi_h + \phi_d) \sim 0.94$ and for PEG at $\phi_d/(\phi_h + \phi_d) \sim 0.15$. Under condition $\phi_d/(\phi_h + \phi_d) \sim 0.15$, the measured scattering will be dominated by scattering from the polymer where contributions from $S_{pp}(q)$ and $S_{pc}(q)$ will be present since $\Delta\rho_c \sim 0$ in the above equation.

Scattering measurements at a fixed $\phi_c$ were made at seven D/H-EtOH ratios corresponding to $\Delta\rho_p \sim 0$, $\Delta\rho_c \sim 0$ and five intermediate values close to $\Delta\rho_c=0$. Changes in the scattering profile with variations in $\phi_h/\phi_d$ are shown in Figure S16 for a silica volume fraction sample of $\phi_c=0.1$ in a solution containing $R_P=0.4$ PEG where only the D/H ratio is varied. At low D-EtOH concentrations ($\phi_d/(\phi_h + \phi_d)\sim0.15$), the scattering is dominated by the particles, which is proportional to $P_c(q)S_{cc}(q)$. As the D-EtOH concentration increases, substantial qualitative changes are observed in the scattering profile. To extract each $S_{ij}(q)$ from above equation, we first determine $\Delta\rho_c$ and $\Delta\rho_p$ from the known $\rho_c$ and $\rho_s$ at each D/H-EtOH ratio using the information in Figure S16. Where $\Delta\rho_c = 0$, above equation is simplified as $I(q) \sim A\Delta\rho_c^2 P_c(q)S_{cc}(q)$. $S_{cc}(q)$ is obtained by dividing the scattering intensity from the concentrated particle suspension by its dilute limit analog at the same $R_P$. After the step, only $S_{pp}(q)$ and $S_{pc}(q)$ remain unknown. At each scattering vector, one then has seven experimental data points at points according to the D/H-EtOH ratios and two unknowns allowing us to minimize uncertainty in the two unknowns at each $q$. We solve $S_{pp}(q)$ and $S_{pc}(q)$ using multiple linear regression fitting method.

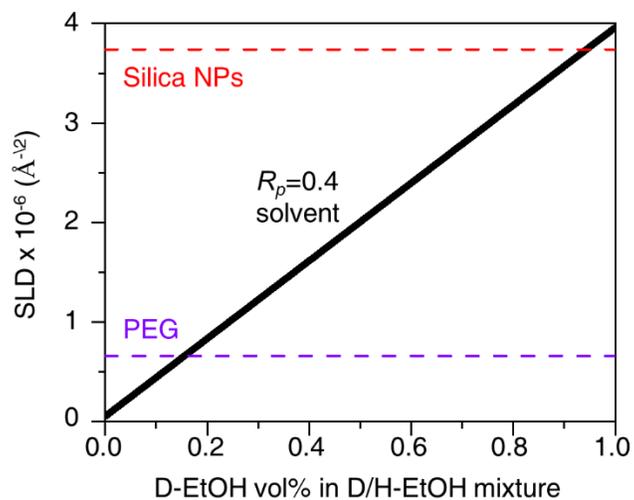

Figure S16. Scattering length density (SLD) as a function of D-EtOH vol% in D/H-EtOH mixture. The theoretical contrast matching points for PEG and silica NPs are ~0.15 and 0.96, respectively.